\documentclass[final,1p,times]{elsarticle}
\usepackage{amssymb}
\usepackage{amsmath}
\usepackage{mathrsfs}
\usepackage{hyperref}
\usepackage{longtable}
\usepackage{xcolor}
\usepackage{xspace}
\usepackage{marginnote}
\usepackage[normalem]{ulem}

\journal{Computers in Biology and Medicine}

\begin{document}
%\linenumbers

\begin{frontmatter}

%% Title, authors and addresses

%% use the tnoteref command within \title for footnotes;
%% use the tnotetext command for theassociated footnote;
%% use the fnref command within \author or \affiliation for footnotes;
%% use the fntext command for theassociated footnote;
%% use the corref command within \author for corresponding author footnotes;
%% use the cortext command for theassociated footnote;
%% use the ead command for the email address,
%% and the form \ead[url] for the home page:
%% \title{Title\tnoteref{label1}}
%% \tnotetext[label1]{}
%% \author{Name\corref{cor1}\fnref{label2}}
%% \ead{email address}
%% \ead[url]{home page}
%% \fntext[label2]{}
%% \cortext[cor1]{}
%% \affiliation{organization={},
%%             addressline={},
%%             city={},
%%             postcode={},
%%             state={},
%%             country={}}
%% \fntext[label3]{}

\title{Vascular Segmentation of Functional Ultrasound Images using Deep Learning}

%% use optional labels to link authors explicitly to addresses:
%% \author[label1,label2]{}
%% \affiliation[label1]{organization={},
%%             addressline={},
%%             city={},
%%             postcode={},
%%             state={},
%%             country={}}
%%
%% \affiliation[label2]{organization={},
%%             addressline={},
%%             city={},
%%             postcode={},
%%             state={},
%%             country={}}

\author[label1,label2]{Hana Sebia} %% Author name
\author[label1,label2]{Thomas Guyet} %% Author name
\author[label2,label3]{Mickaël Pereira}
\author[label4]{Marco Valdebenito}
\author[label1,label2]{Hugues Berry} %% Author name
\author[label2,label3,label4]{Benjamin Vidal} %% Author name

%% Author affiliation
\affiliation[label1]{organization={AIstroSight, Inria, Hospices Civils de Lyon},%Department and Organization
            city={Villeurbanne},
            %postcode={69603}, 
            %state={},
            country={France}}
%% Author affiliation
\affiliation[label2]{organization={University Claude Bernard Lyon 1},%Department and Organization
            city={Villeurbanne},
            %postcode={69603}, 
            %state={},
            country={France}}

\affiliation[label3]{oragnization={Lyon Neuroscience Research Center},
            city={Lyon},
            country={France}}

\affiliation[label4]{oragnization={CERMEP-Imaging Platform},
            city={Lyon},
            country={France}}

%% Abstract
\begin{abstract}
Segmentation of medical images is a fundamental task with numerous applications. While MRI, CT, and PET modalities have significantly benefited from deep learning segmentation techniques, more recent modalities, like functional ultrasound (fUS), have seen limited progress.
fUS is a non invasive imaging method that measures changes in cerebral blood volume (CBV) with high spatio-temporal resolution.
However, distinguishing arterioles from venules in fUS is challenging due to opposing blood flow directions within the same pixel.
Ultrasound localization microscopy (ULM) can enhance resolution by tracking microbubble contrast agents but is invasive, and lacks dynamic CBV quantification.
In this paper, we introduce the first deep learning-based application for fUS image segmentation, capable of differentiating signals based on vertical flow direction (upward vs. downward), using ULM-based automatic annotation, and enabling dynamic CBV quantification. In the cortical vasculature, this distinction in flow direction provides a proxy for differentiating arteries from veins.
We evaluate various UNet architectures on fUS images of rat brains, achieving competitive segmentation performance, with 90\% accuracy, a 71\% F1 score, and an IoU of 0.59, using only 100 temporal frames from a fUS stack.
These results are comparable to those from tubular structure segmentation in other imaging modalities.
Additionally, models trained on resting-state data generalize well to images captured during visual stimulation, highlighting robustness. 
Although it does not reach the full granularity of ULM, the proposed method provides a practical, non-invasive and cost-effective solution for inferring flow direction—particularly valuable in scenarios where ULM is not available or feasible.
Our pipeline shows high linear correlation coefficients between signals from predicted and actual compartments,
showcasing its ability to accurately capture blood flow dynamics.

\end{abstract}

%%Graphical abstract
%\begin{graphicalabstract}
%\includegraphics[width=\linewidth]{avseg_schema_3.pdf}
%\end{graphicalabstract}

%%Research highlights
%\begin{highlights}
%\item We present the first deep learning tool specifically designed for vascular segmentation in functional ultrasound imaging.
%\item Our pipeline demonstrates competitive segmentation performance compared to state-of-the-art methods (90\% accuracy and 71\% F1 score).
%\item This tool improves the interpretation of fUS power Doppler signals, supporting preclinical neurovascular research.
%\item Our models exhibit strong generalization across varying brain states, including during visual stimulation.
%\item Our models achieve optimal performance with only 100 temporal frames from the fUS stack.

%\end{highlights}

%% Keywords
\begin{keyword}
functional ultrasound \sep segmentation \sep ultrafast ultrasound localization microscopy\sep preclinical \sep medical images \sep neuroscience
%% keywords here, in the form: keyword \sep keyword

%% PACS codes here, in the form: \PACS code \sep code
\PACS 43.35.Yb %Ultrasonic instrumentation and measurement techniques 
\sep 87.63.dk %Doppler imaging in ultrasonography, 
\sep 87.19.La %neuroscience

%% MSC codes here, in the form: \MSC code \sep code
%% or \MSC[2008] code \sep code (2000 is the default)

\MSC[2008] 62H35 %Image analysis
\sep 68T05 % AI/learning
\sep 82C32 %: Neural nets

\end{keyword}

\end{frontmatter}

%% Add \usepackage{lineno} before \begin{document} and uncomment 
%% following line to enable line numbers
%% \linenumbers

%% main text
%%

%% Use \section commands to start a section
\section{Introduction}
\label{sec1}
Blood flow and volume are closely linked to neuronal activity changes in the brain, a phenomenon known as neurovascular coupling. This relationship has enabled functional neuroimaging techniques, such as functional ultrasound imaging, to successfully probe brain activity at the mesoscopic scale in both animals and humans. 
Functional ultrasound imaging (fUS) is a neuroimaging technique that leverages hemodynamic signals to map brain activity, achieving high spatial resolution down to 100 µm and temporal resolution as low as 400 ms~\cite{mace2011functional, deffieux2018functional, deffieux2021functional}. This method captures changes in cerebral blood volume related to neuronal activity through ultra-fast Power Doppler imaging. These images are analyzed to understand brain functions and pathologies~\cite{vidal2020functional, droguerre2022impaired, vidal2022pharmaco}.
Accurately identifying whether signals originate from arterioles or venules in these images is crucial to improve our understanding of brain vascular dynamics, as these distinct vascular compartments are differently involved in spontaneous fluctuations of vessel tone and functional hyperemia~\cite{Chen2023.11.07.566048, he2018ultra, renaudin2022functional, bourquin2021vivo} as well as energy supply~\cite{qi2021control} and  waste clearance~\cite{iliff2013cerebral}.
Unfortunately, while the spatial resolution of fUS is relatively high for in vivo mesoscopic brain imaging compared to fMRI, it is insufficient to resolve individual blood vessels. 
Although Color Doppler sequences in ultrasound imaging can determine blood flow direction --~which helps in differentiating opposing flows in cortical arterioles and venules~-- this method still struggles to clearly identify these compartments.
This limitation arises because upward and downward flows often coexist within the same 100-micron voxels, canceling each other out and obscuring clear compartmental distinction.
Additionally, in commercially available fUS setups, the raw data necessary for speckle tracking or color flow doppler is not accessible to the user.
These setups provide only Power Doppler image stacks, which are generated through an entirely different processing workflow. 
Even in cases where such raw data are available, their substantial volume would present significant computational challenges, making the processing and extraction of this information highly demanding in terms of resources.

To overcome this, ultrasound localization microscopy (ULM) is an advanced imaging modality that also employs ultra-fast sampling techniques to significantly enhance spatial resolution to just a few microns~\cite{errico2015ultrafast, couture2018ultrasound}. 
This advanced approach not only improves spatial resolution but also enables the determination of blood flow direction. 
Nonetheless, it requires the intravenous injection of micro-bubble contrast agents and the tracking of their position along a protracted time period to create a detailed vascular map. This map differentiates between upward and downward flows based on the movement direction of the micro-bubbles.
While ULM provides comprehensive insights into cerebral blood flow dynamics, it lacks the temporal resolution of functional ultrasound. Moreover, its deployment is complex and invasive, necessitating the injection of contrast agents and presenting significant challenges in managing large data volumes required for image reconstruction.

This study leverages deep learning to enhance fUS image analysis by circumventing the conventional use of ULM.
Specifically, we aimed to differentiate between upward and downward blood flows, corresponding to arterial and venous structures in the cortical region, directly from fUS images, without relying on ULM or requiring a particular data processing of the intermediate raw data.
Our approach focused on automating an image segmentation task by training a deep learning model on a dataset specifically annotated to identify these vascular compartments.
The process of constructing annotations, based on available ULM images, is detailed in Section~\ref{masks_construction}. 
Once the model is trained, it can be applied to new fUS images to accurately retrieve vascular structures without requiring ULM sessions or manual annotation.
The challenge was to identify a deep learning architecture that could achieve high accuracy. We focused our investigation on UNet architectures, known for their effectiveness on medical images, even when trained on small datasets~\cite{azad_medical_2022}.
To address this, we proposed a benchmarking module that compares multiple segmentation architectures as detailed in Section~\ref{model_choice}.
Our results, summarised in Section~\ref{sec:exp:comparison}, demonstrated a segmentation quality comparable to that achieved for other imaging modalities (MRI, OCT,...) on vascular and tubular structures. 
Additionally, we investigated, in Section~\ref{sec:exp:stackdepth}, how incorporating different temporal frames from fUS data affects the segmentation quality, exploring potential improvements in accuracy. Finally, we showed, in Section~\ref{sec:exp:crosscondition}, that training a model on fUS images acquired in a resting state generalize to images acquired under visual stimulation with the same precision, without requiring re-training. 
In Section~\ref{sec:exp:visu}, we illustrated how this segmentation approach can be used in a practical way to improve fUS signal interpretation.

\section{Related work}
\label{sec2}

\subsection{Different applications of Deep Learning on functional ultrasound}
\label{subsec1}
Despite the potential benefits, fUS has seen few studies that make use of advanced computational methods, particularly for image segmentation task. The existing research primarily focuses on other aspects, such as improving image reconstruction from sparse data and enhancing sensitivity in neuroimaging applications.
For instance, Di Ianni and Airan~\cite{9701931} have developed a deep learning platform known as Deep-fUS, which significantly reduces the amount of ultrasound data required while maintaining image quality. Their work demonstrates the potential for deep learning to significantly streamline fUS imaging using portable devices and resource-limited settings. This approach also highlights the innovative use of CNN in reconstructing power Doppler images from under-sampled sequences.
Similarly to fUS, ULM has also seen limited exploration with deep learning. 
van Sloun et al.~\cite{9257449}~have leveraged CNN to achieve super-resolution in ULM, significantly improving the precision of micro-bubble localization, while Milecki et al.~\cite{9345725} have developed a spatio-temporal framework that refines the process of micro-bubble tracking. 
While the aforementioned works have made substantial advancements in improving the quality of images produced by fUS and ULM, they primarily focus on enhancing image reconstruction and resolution.

\subsection{UNets for medical image segmentation}
\label{sec:subsec2}
Medical image segmentation has witnessed major advancements, particularly through the use of deep learning models~\cite{medical_image_seg_dl_survey, wang_medical_2022}. The UNet architecture has demonstrated remarkable success, providing precise segmentation across diverse imaging modalities such as Ultrasound~\cite{cai_ultrasound_2023}, PET~\cite{bourigault2021multimodal}, MRI~\cite{dolz2018ivd} and CT~\cite{li2018h}. 
Its design, characterized by a symmetric encoder-decoder structure and extensive use of skip connections~\cite{10230609,peng_u-net_2023}, facilitates significantly the segmentation of various organs and tumors such as skin lesions~\cite{tong_ascu-net_2021}, liver and rectal tumors in CT and MRI scans~\cite{ayalew_modified_2021,zhang_imaging_2024}, or breast tumors in ultrasound imaging~\cite{pramanik_dbu-net_2023}.
Enhancements to UNet have also adapted it for volumetric data from MRI and CT, extending its structure to better analyze spatial relationships in 3D space~\cite{qamar2020variant, milletari2016v, raza2023dresu}, which is essential for organs and vascular segmentation~\cite{huang2018robust}. 
Numerous reviews continue to discuss these advancements and their contributions to the field~\cite{liu_deep_2019,azad_medical_2022,rayed_deep_2024}.

\subsection{Vessel segmentation}
\label{subsec3}
Medical imaging segmentation has also expanded to encompass more complex and subtle structures, such as blood vessels. This task has primarily focused on the segmentation of cerebral and retinal vessels, crucial for diagnosing and monitoring neurological and ocular diseases~\cite{goni_brain_2022, chen_retinal_2021}. 
This field is broadly categorized into two main areas: 2D and 3D segmentation. Volumetric imaging like Magnetic Resonance Angiography~(MRA) allows for constructing detailed 3D vascular tree structures, with recent advances using enhanced UNet architectures~\cite{wu2023transformer,yang2024segmentation}. However, given the specific task we aim to address, our research focuses particularly on 2D segmentation. This emphasis on 2D analysis led us to explore domain-specific optimizations that significantly improve segmentation outcomes. Notably, Shit et al.~\cite{shit_cldice_2021}~introduced the $centerlineDice$ metric, which measures similarity across segmentation masks and their morphological skeletons, ensuring topological continuity. Furthermore, Zhou et al.~\cite{ZHOU2024103098}~has advanced this area by embedding vessel density and fractal dimensions directly into the loss function, enabling refined multi-class vessel segmentation that categorizes pixels into veins, arteries, background or uncertain regions. This novel regularization has also improved biomarker quantification, directly contributing to downstream clinical tasks.

%%%%%%%%%%%%%%%%%%%%%%%%%%%%%%%%%%%%%%%%%%%%%%%%%%%%
%%%%%%%%%%%%%%%%%%%%%%%%%%%%%%%%%%%%%%%%%%%%%%%%%%%%
\section{Method}
\label{sec3}

\subsection{fUS acquisition}
For the acquisition of functional ultrasound imaging, we performed experiments on the brain of $35$ anesthetized rats. 
A few days before imaging, the rats underwent a skull-thinning procedure to enhance the signal-to-noise ratio, as described in~\cite{vidal2020functional}.
The rats were anesthetized using a combination of subcutaneous medetomidine and isoflurane~\cite{grandjean2023consensus}.
Imaging was performed using a preclinical ultrasound imaging system (Iconeus V1, Paris, France).  Doppler vascular images were obtained using the Ultrafast Compound Doppler Imaging technique~\cite{bercoff2011ultrafast}. Each frame corresponded to a Compound Plane Wave frame~\cite{montaldo2009coherent} resulting from the coherent summation of backscattered echoes obtained after successive tilted plane waves emissions. 
Stacks % of 200 compounded frames, 
acquired at a $500~Hz$ frame rate, were processed with a dedicated spatiotemporal filter based on Singular Value Decomposition (SVD)~\cite{demene2015spatiotemporal} to isolate the blood volume signal from tissue signals, thus generating Power Doppler images.
These final images have a temporal resolution of $0.4~s$, a spatial resolution of $100$ ~\textmu m and are directly proportional to the cerebral blood volume.

In the initial phase, we conducted a 20-minute resting-state recording.
This generated a 2D + time image stack consisting of $3\,000$~frames, each with a spatial resolution of $112\times128$ pixels (100 microns/pixel). This first stack details the cerebral activity signals associated with hemodynamic changes at rest.
Following the initial data collection, a second acquisition was conducted immediately. 
During this phase, we implemented visual stimulation by maintaining the anaesthetised rat in the dark and exposing it to periodic light flashes, as described in~\cite{droguerre2022impaired}.
Stimulation runs consisted of black and white flickering on a screen placed in front of the animal (flickering at a frequency of $3~Hz$ and continuous black screen for rest). The stimulation pattern consisted in $30~s$ of initial rest followed by runs of $30~s$ of flicker and $45~s$ of rest, repeated four times for a total duration of $330~s$. 
This session was shorter than the first, yielding an image stack of $825$ frames at the same spatial resolution and acquisition frequency. This second stack captures therefore cerebral activity linked to hemodynamic changes during visual stimulation.

\subsection{ULM acquisition}
\begin{figure}[t]
\centering
\includegraphics[width=\linewidth]{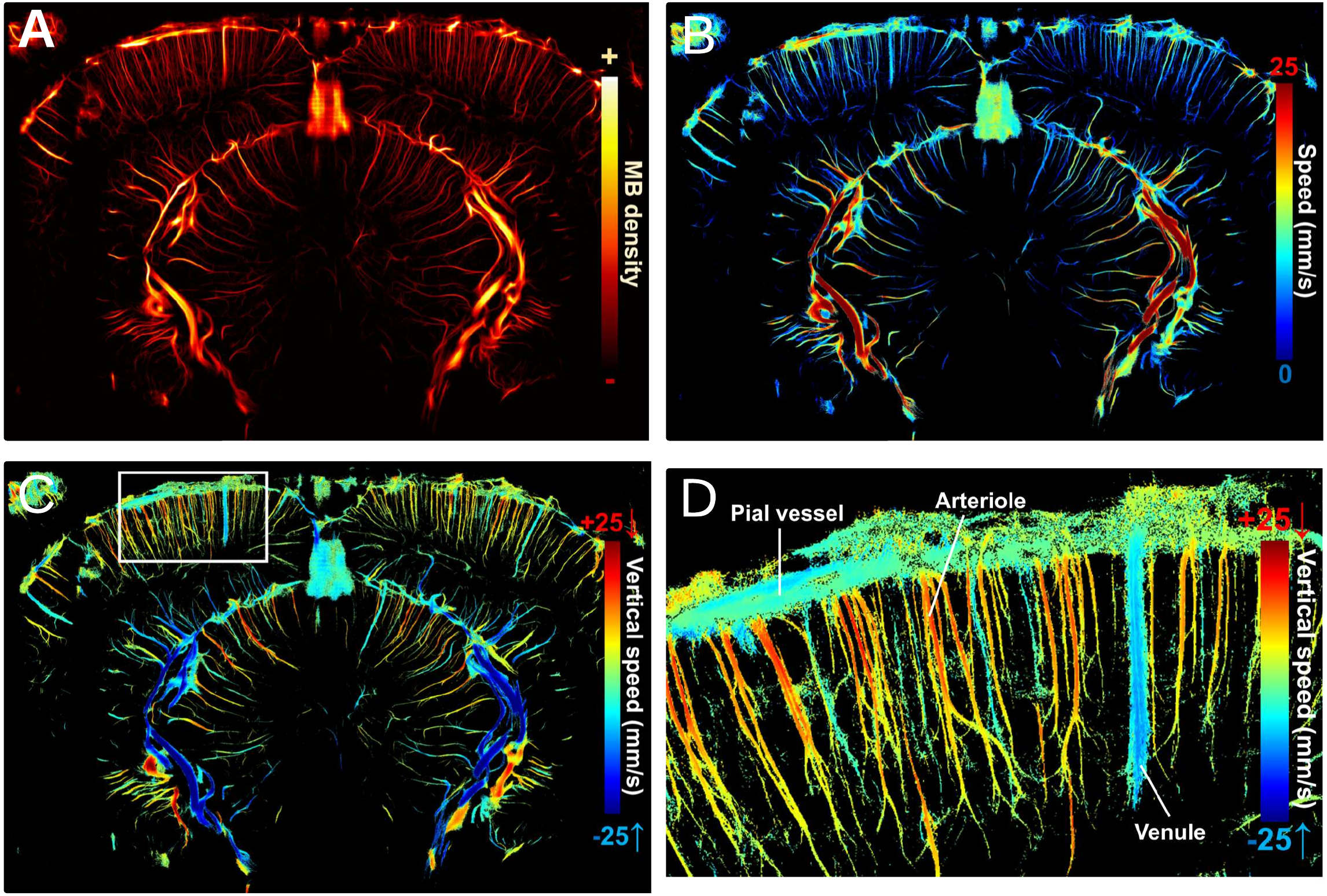}
\caption{ULM modalities: (A) micro-bubble density, (B) micro-bubble velocity, (C) micro-bubble velocity over the Z-axis, (D) Zoom over arterioles and venules from micro-bubble velocity following the Z-axis.} \label{fig:ulm}
\end{figure}

Immediately following the initial fUS scans, we conducted an ultrasound localization microscopy session using the same equipment (Iconeus V1). 
For this, we injected contrast-enhancing micro-bubbles into the bloodstream of the rat via the tail vein and performed imaging over $5$ minutes. 
The raw data collected using a dedicated Iconeus software, based on the methodology described by Errico et al.~\cite{errico2015ultrafast}, were then used to construct three ultra-high resolution imaging modalities. 
These images, illustrated in Figure~\ref{fig:ulm}, detail the cerebral vascular tree at a fine scale with different types of information: (A) micro-bubble density, (B) velocity, and (C) axial velocity along the Z-axis. 
The resolution is 2 microns per pixel ($7\,011\times5\,494$ pixels) for micro-bubble density and 10 microns per pixel ($7\,008\times5\,490$ pixels) for velocity. The lower resolution for velocity results from applying a filter to remove outliers, based on the individual tracking of micro-bubbles.
The construction of these modalities is performed by the Iconeus software and requires approximately $6$~hours of computation per experiment.
 
\subsection{fUS Automatic Annotation}
\label{masks_construction}
The above experiments provided us with two stacks of fUS (resting state and visual stimulation), paired with one ULM image for each animal. 
We used the ULM image for the vascular annotation within the fUS stack, as it provides the necessary information, particularly in the cortical region, which is the primary focus of our fUS analysis.
More specifically, the ULM Z-axis velocity modality is used to automatically distinguish different vascular compartments: in the cortex, positive velocity values, associated with downward flow, indicate the movement of micro-bubbles through arterioles, while negative values, associated with upward flow, reflect their movement through venules.
This assumption applies primarily to the cortical region, where this relationship can be easily made when the vessels are vertically oriented. 
We acknowledge that it may not apply in other regions.
The choice of using velocity along the z-direction was motivated by the predominant vertical orientation of blood vessels in the central portion of the cortex, where most of our analysis was focused. While this assumption does not capture all vascular geometries, and branching vessels in lateral areas or oblique orientations could be misclassified, it was chosen as a practical approach given the constraints of the data and the need for an automated annotation method.

\begin{figure}[t]
\centering%
\includegraphics[width=\linewidth]{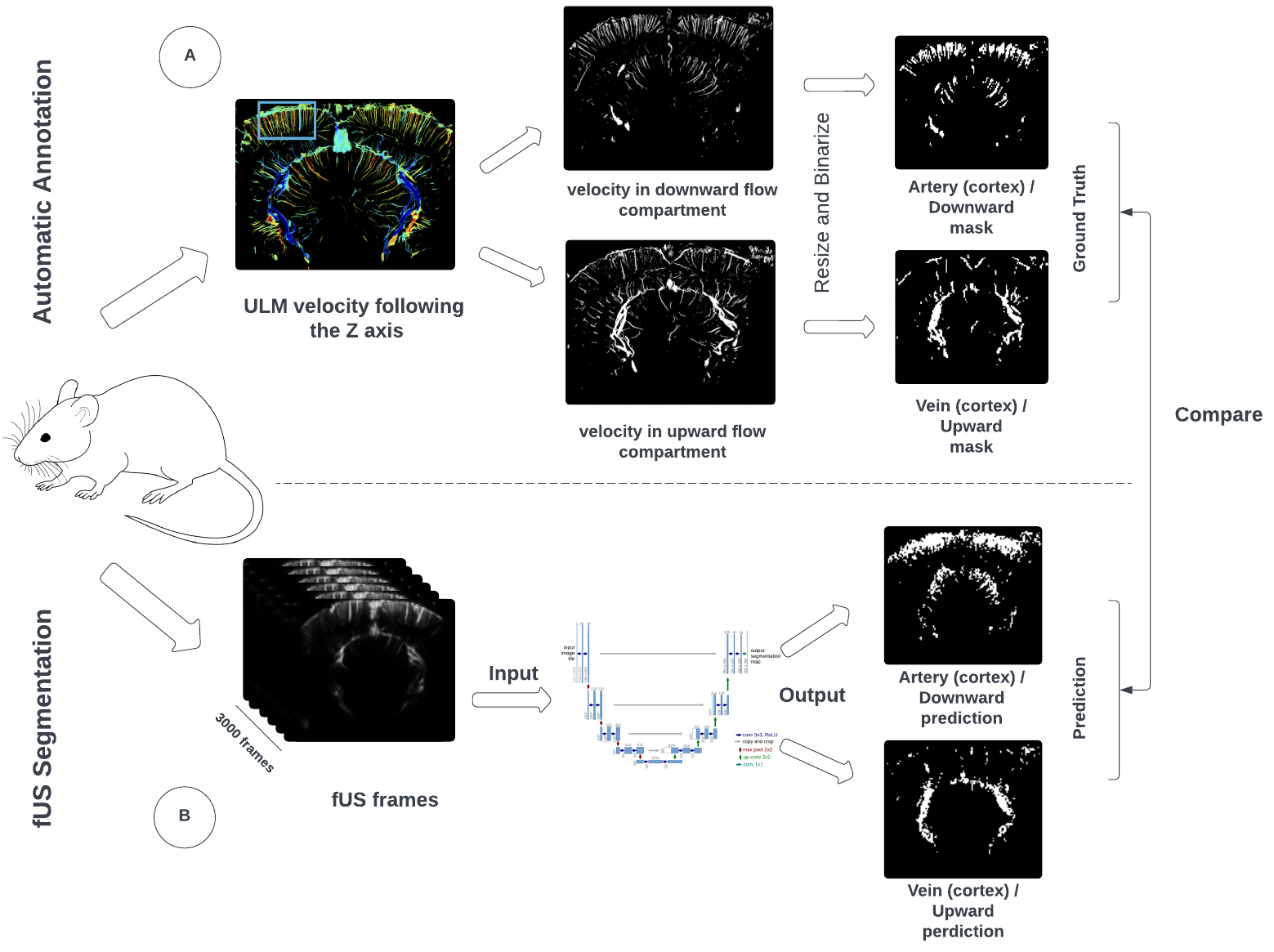}
\caption{(A) Process of constructing fUS annotations --~downward and upward masks~-- from ULM images. (B) Segmentation framework of fUS stacks based on UNet architecture.} 
\label{fig:avseg_schema}
\end{figure}

To automatically differentiate blood flow direction, we applied thresholding to the Z-axis velocity image, as illustrated in Figure~\ref{fig:avseg_schema}-A. 
The images were then resized using bilinear interpolation to match the spatial resolution of the fUS stack. 
Since our primary focus is on identifying vascular structures rather than quantifying velocity, the resulting images were converted into binary masks. 
To ensure robust annotations, we set a conservative threshold of $0.05$ to exclude ultra-fine structures that fUS is unlikely to detect, thereby producing more accurate and useful vascular masks.
These masks served as ground truth annotations for both resting and visual stimulation states, as they represent the anatomical structure of the brain vascular tree and do not contain functional information.
It is important to note that resizing ultra-high resolution images introduced a small proportion of mixed pixels belonging to both downward and upward masks. However, since these mixed pixels constituted less than 1\%, they were retained to avoid artifacts that might arise from selective pixel exclusion.

\subsection{Segmentation framework}
We approached our problem of automatic compartmentalization of fUS imaging as a multi-class segmentation task where each pixel in the fUS image must be classified as upward flow, downward flow or background. To this end, we trained a deep neural network to generate upward and downward masks for each fUS image stack input. These generated masks are then compared to the ground truth derived from ULM to adjust the weights of the network. This pipeline is detailed in Figure~\ref{fig:avseg_schema}-B.
We opted for a UNet architecture and some of its variants for the segmentation. Additionally, we evaluated various loss strategies to optimize feature and pixel-level classification. Our aim was to determine the most effective model and the most appropriate loss function for accurate fUS image segmentation.
Once trained, the model is capable of generating masks based on new fUS images. 
Those masks are then projected on the fUS stack, enabling the visualization of cerebral blood volume evolution along the temporal axis while distinguishing the concerned compartments.

\subsubsection{Selected models}
\label{model_choice}
As discussed in Section~\ref{sec:subsec2}, there have been significant advances and enhancements in the UNet architecture. We selected a benchmark reviewed in~\cite{azad_medical_2022} from their Awesome-U-Net\footnote{\url{https://github.com/NITR098/Awesome-U-Net/tree/main}} GitHub repository. 
In addition to the classic UNet architecture~\cite{zhou_unet_2018}, known for its robustness with limited datasets and its suitability for medical imaging applications, six architectures were chosen for their ability to address various complexities encountered in medical image segmentation:
\begin{itemize}

    \item \textbf{Residual Net (2016)}~\cite{drozdzal_importance_2016} combines the strengths of UNet with ResNet, allowing for deeper networks without the risk of vanishing gradients. It uses recurrent convolutions for improved feature representation.

    \item \textbf{UNet++ (2018)}~\cite{zhou_unet_2018} introduces nested, dense skip pathways. These additional connections help in better feature propagation across the network, reducing the semantic gap between the feature maps of the encoder and decoder stages.

     \item \textbf{Attention UNet (2018)}~\cite{oktay2018attention}, by incorporating attention gates, selectively emphasizes important features and suppresses irrelevant neuron activation, which is crucial in medical images where the focus area is often surrounded by a large amount of noise.
    
    \item \textbf{MultiresNet (2020)}~\cite{ibtehaz_multiresunet_2020} employs a multi-resolution convolutional approach to handle diverse image scales effectively within the segmentation process. It integrates layers that process multiple scales simultaneously, which helps in capturing both global and local contextual details more effectively.

    \item \textbf{UCTransNet (2022)}~\cite{UCTransNet} proposes a Channel Transformer module (CTrans) with attention mechanism that replaces the skip connections in original U-Net. Specifically, the CTrans guides the fused multi-scale channel-wise information to effectively connect to the decoder features for eliminating the ambiguity.

    \item \textbf{TransUNet (2024)}~\cite{chen2024transunet} encapsulates self-attention into two key modules: (1) a Transformer encoder tokenizing image patches from a CNN feature map, facilitating global context extraction, and (2) a Transformer decoder refining candidate regions through cross-attention between proposals and U-Net features.

\end{itemize}

The selected architectures are originally designed to process 2D images as input and produce 2D segmentations, except for TransUNet, which also includes a 3D implementation. 
However, to leverage the temporal richness of our fUS data, we opted to 
feed the UNet with multiple frames from a stack as many input channels. 
This approach allowed us to exploit the inter-frame dynamics while preserving the 2D structure of the existing architectures, avoiding the added complexity of adapting the networks. This trade-off ensures optimal capture of spatio-temporal information while maintaining the efficiency of the pre-existing models.

\subsubsection{Selected losses}
 In this section, $S\in[0,1]^P$ represents the segmentation mask produced by the model from fUS image containing $P=w\times h$ pixels ($h$ and $w$ are respectively the height and the width of the image).
The ground truth image, denoted $T$, consists of a ternary mask derived from upward and downward masks. Specifically, when needed, it is expressed as  $T_c\in\{0,1\}^P$ where $ c \in \{u,d,b\}$~-- with $u$ indicating upward flow compartment, $d$ indicating downward flow compartment, and $b$ representing the background. $S$ can be specified in a similar manner when necessary.
In the remaining of this section, $S_{i,c}$ (resp. $T_{i,c}$) is the predicted probability (resp. ground truth label) for class $c$ at pixel $i$
%The dimensions of the image are given by the height $h$ and width $w$.

For the first loss function, we proposed to use a combination of Cross-Entropy and Dice loss. Cross-entropy is effective for pixel-wise segmentation tasks. For each pixel \( i \), the Cross-Entropy Loss is defined as:
\begin{equation}
    %\text{CE}_i(S, T) = - \sum_{c=1}^{C} T_{i,c} \log(S_{i,c})
    \text{CE}_i(S, T) = - \sum_{c\in\{u,d,b\}} T_{i,c} \log(S_{i,c}).
\end{equation}

This approach penalizes incorrect classifications at the pixel level.
To obtain the final Cross-Entropy term, the $\text{CE}_i$ is averaged over all pixels in the image:
\begin{equation}
    \mathscr{L}_{CE}(S, T)= 
    \frac{1}{P} \sum_{i=1}^{P} \text{CE}_i\text{(S, T)}.
\end{equation}

Dice loss is particularly useful for addressing class imbalances by focusing on the overlap between the predicted and ground truth masks. This is especially important in our case, as we have a significant class imbalance with far more background pixels compared to upward and downward pixels.
Specifically, the Dice coefficient for a class $c\in\{u,d,b\}$ is defined as follows:
\begin{equation}
    \text{Dice}_c\text{(S, T)} = \frac{2 \sum_i S_{i,c} T_{i,c}}{\sum_i S_{i,c} + \sum_i T_{i,c} + \varepsilon}
\end{equation}

where \( \varepsilon \) is a small constant added to avoid division by zero. 
The Dice Loss for each class is then calculated as:
\begin{equation}
    \mathscr{L}_{Dice}(S, T, c)
    = 1 - \text{Dice}_c\text{(S, T)}
\end{equation}

To obtain the final Dice term, we compute the Dice Loss for each class and then take the average across all classes \( C \):
\begin{equation}
    \mathscr{L}_{Dice}(S, T)= \frac{1}{C} \sum_{c\in\{u,d,b\}} \mathscr{L}_{Dice}(S, T, c)
\end{equation}

The final formulation of the combined loss function is given by a weighted sum of the previously detailed terms, as follows:
\begin{equation}
    \mathscr{L}_{Dice\_CE}(S, T)= \alpha\, \mathscr{L}_{CE}(S, T) + \beta\, \mathscr{L}_{Dice}(S, T), \quad 
    \alpha, \beta \in [0,1]
\end{equation}

where \( \alpha \) and \( \beta \) are weights that balance the contribution of each loss term, and they are not required to sum to 1.

For the next loss functions, in combination with Cross-Entropy, we incorporate a regularization approach introduced in~\cite{ZHOU2024103098}. This approach includes two key terms: vessel density and fractal dimension. The vessel density measures a ratio of vessel area to a whole area as follows:

\begin{equation*}
\label{vessel_density}
    \mathscr{L}_{V}(S, T)= \displaystyle\left\lvert \frac{\sum_{i=1}^{P} S_{i,d} - \sum_{i=1}^{P} T_{i,d}}{h \times w} \right\rvert + \displaystyle\left\lvert \frac{\sum_{i=1}^{P} S_{i,u} - \sum_{i=1}^{P} T_{i,u}}{h \times w} \right\rvert
\end{equation*}

To determine the fractal dimension $d_f$ of vessels, we rely on the Minkowski-Bouligand dimension~\cite{falconer2014fractal}, commonly referred to as the box-counting dimension. This measure evaluates the complexity of vessel morphology by covering the image with a grid of boxes of a given size and counting how many boxes contain a part of the vessel structure. 
As the box size $\varepsilon$ decreases, more boxes are required to cover the vessels, and the scaling behavior $N\propto \varepsilon^{d_f}$ with $N$ the number of required boxes, defines the fractal dimension $d_f$.
More explicitly, we set the box sizes $\varepsilon=\left \{2^i | i\in \mathbb{Z}, 2\leq 2^i\leq \min \left\{h,w\right\}\right\}$ and count the box number $N_S(\varepsilon)$ needed to cover the segmentation mask and $N_T(\varepsilon)$ to cover ground truth mask. The loss is then calculated as:
\begin{equation*}
\mathscr{L}_{B}(S,T)= \sum_{c\in\{u,d,b\}}\frac{1}{\sqrt{\sum_{i=1}^{P}\varepsilon_i^2}}\cdot \sum_{i=1}^{P}\sqrt{\varepsilon_i\cdot  \left(\frac{N_{T_c}(\varepsilon_i)-N_{S_c}(\varepsilon_i)}{N_{T_c}(\varepsilon_i)}\right)^2}
\end{equation*}

$\varepsilon_i$ was empirically configured to weight more on the error of large-size box.

To achieve accurate multi-class segmentation masks, we construct the $\mathscr{L}_{CF}$ Loss (Clinically-relevant Feature-optimised Loss) by combining the feature-based loss functions, $\mathscr{L}_{V}$ and $\mathscr{L}_{B}$, with the pixel-wise Cross-Entropy Loss. Following the approach proposed in~\cite{ZHOU2024103098}, we formulate the $\mathscr{L}_{CF}$ Loss in three configurations to balance clinical relevance with segmentation accuracy:

\begin{align}
    &\mathscr{L}_{CF\_B}(S,T) =  \alpha \, \mathscr{L}_{CE}(S, T) + \gamma \, \mathscr{L}_B(S,T) \\
    &\mathscr{L}_{CF\_V}(S,T) =  \alpha \, \mathscr{L}_{CE}(S, T) + \beta \, \mathscr{L}_V(S,T) \\
    &\mathscr{L}_{CF}(S,T) =  \alpha \, \mathscr{L}_{CE}(S, T) + \gamma \, \mathscr{L}_B(S,T) + \beta \, \mathscr{L}_V(S,T)
\end{align}

where $ \alpha$, $\beta$ and $\gamma$ are the loss weights for Cross-Entropy, vessel density and box count terms respectively. The values used for these weights are given in Section~\ref{sec:hyperparameter-setup} below.

%%%%%%%%%%%%%%%%%%%%%%%%%%%%%%%%%%%%%%%%%%%%%%
%%%%%%%%%%%%%%%%%%%%%%%%%%%%%%%%%%%%%%%%%%%%%%
\section{Experiments}
\label{sec:experiments}

\subsection{Dataset}

As mentioned in the Method Section, our data were derived from experiments on 35 different rats, providing us with 35 fUS stacks in both resting state and under visual stimulation, annotated as described in Section~\ref{masks_construction}. 
No pre-processing was applied to these fUS images, except for the UCTransNet and TransUNet models, where they were resized, as well as the constructed masks, to 128x128 pixels to meet the requirement of square input dimensions.
To improve the robustness and generalizability of our models, we employed data augmentation techniques, including random horizontal and vertical flips and random rotations. This augmentation was applied in real-time during the training phase, where each frame of the input stack was modified with a random transformation at each training iteration. The corresponding ULM masks were augmented with the same transformations to maintain precise alignment with the fUS frames. To ensure accuracy in rotation, these flips and rotations were performed after resizing.
Finally, we conducted a 7-fold cross-validation to ensure the reliability of our results, using data from 30 experiments randomly chosen for training and reserving 5 for testing.

\subsection{Evaluation metrics}

To provide a comprehensive evaluation of the segmentation performance, we selected the metrics detailed above:
\begin{equation*}
\text{Accuracy} = \frac{TP + TN}{TP + TN + FP + FN}, \quad \text{Specificity} = \frac{TN}{TN + FP}
\end{equation*}

\begin{equation*}
\text{Precision} = \frac{TP}{TP + FP}, \quad \text{Recall} = \frac{TP}{TP + FN}, \quad F1 = \frac{ 2 \times \text{Precision} \times \text{Recall}}{ \text{Precision} + \text{Recall}}
\end{equation*}

In this multi-class setting, $TP$ (True Positives),  $TN$ (True Negatives), $FP$ (False Positives), and $FN$ (False Negatives) denote the counts for each pixel classification category. 

In addition, we also used the Jaccard Index, also known as Intersection over Union (IoU). It is calculated for each class \( c \) individually, and then averaged to provide an overall score:
\begin{equation*}
\text{Jaccard}_c = \frac{|T_c \cap S_c|}{|T_c \cup S_c|}, \quad
\text{Jaccard Index} = \frac{1}{C} \sum_{c=1}^{C} \text{Jaccard}_c
\end{equation*}

$T$ and $S$ denote the ground truth and segmentation masks respectively, as introduced before. 

\subsection{Hyperparameter Setup}
\label{sec:hyperparameter-setup}
For the configuration of the selected deep learning models, we used the hyperparameters recommended in the GitHub repository from which the models were sourced. The only changes made were to the input and output channels; the output was consistently set to three to predict a ternary mask. For the input, as mentioned in Section~\ref{model_choice}, we treated the temporal dimension of the fUS stack as an image channel, setting it to $3\,000$ or $825$ depending on the training set (resting state or under visual stimulation respectively). We chose weights for different losses as follows: $\alpha$ and $\beta$ at $0.5$ in $\mathscr{L}_{Dice\_CE}$,
$\alpha$ at 1 for all $CF$ losses,
$\beta$ and $\gamma$ at $1$ in $\mathscr{L}_{CF\_B}$ and $\mathscr{L}_{CF\_V}$ respectively, and $\gamma$ set at $0.5$ for $\mathscr{L}_{CF}$ as per the default settings in the reference paper. After various trials and errors, we settled on using the Adam optimizer with a learning rate of $10^{-3}$ across $200$ epochs for all training sessions.

\subsection{Results \& Discussion}

\subsubsection{Comparison among models and losses}
\label{sec:exp:comparison}
In this experiment, we trained the selected models using the selected four loss functions on the resting state dataset, with the aim of identifying the most efficient model and the most appropriate loss function. We reported the average metrics across a $7$-fold cross-validation in Table~\ref{tab:comprehensive_model_performance_3000}, page~\pageref{tab:comprehensive_model_performance_3000}. 

The highest performance metrics achieved were an accuracy of $89\%$, an F1 Score of $70\%$, and a Jaccard Index (IoU) of $0.57$, aligning with results typically seen in vessel segmentation~\cite{ZHOU2024103098, shit_cldice_2021}. The best results on average are predominantly achieved by the TransUNet, Attention UNet and UNet++, which seem better suited for this segmentation task due to their ability to focus on relevant regions within the image through attention mechanism. The UCTransNet, UNet and ResNet performed slightly worse yet remained competitive, whereas the MultiResNet was noticeably less effective. This can be explained by its inability to effectively capture the fine details due to its more complex and less targeted architecture.
In terms of loss functions, $CF$ generally outperformed $DICE\_CE$ for Attention UNet, UNet++, and UNet, with negligible differences among the CF variants, suggesting that both fractal dimension and vessel density allowed the models to better identify vascular structures and distinguish vessel boundaries, improving segmentation accuracy.

\begin{figure}[t]
\centering%
\includegraphics[width=\linewidth]{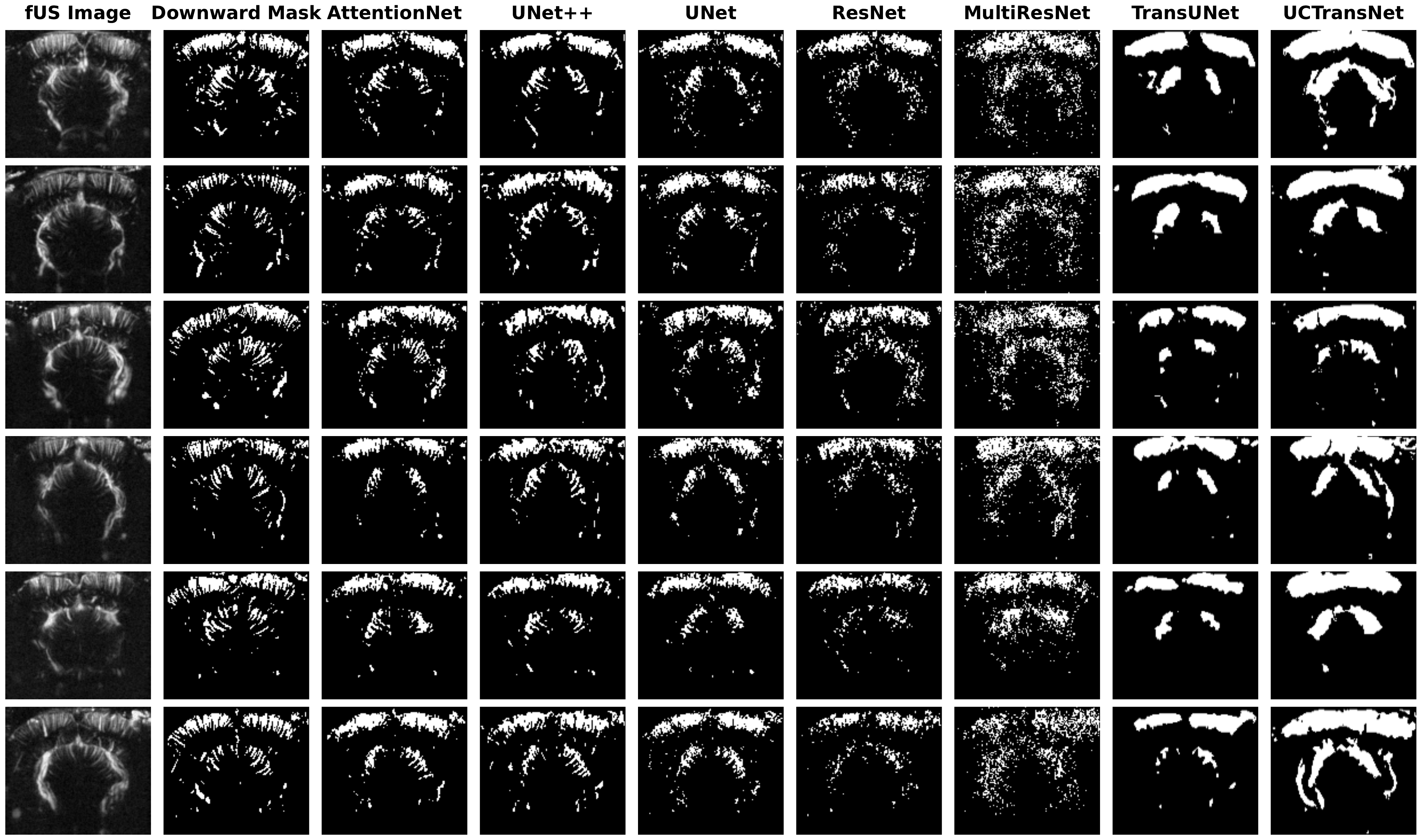}
\caption{Downward prediction examples using $\mathscr{L}_{CF}$. Each row depicts a different image, starting with the first frame of the fUS stack in a log scale, followed by the constructed downward mask from ULM, and the segmentation proposed by the five benchmarked models.}
\label{fig:prediction_artery}
\end{figure}

\begin{figure}[t]
\centering%
\includegraphics[width=\linewidth]{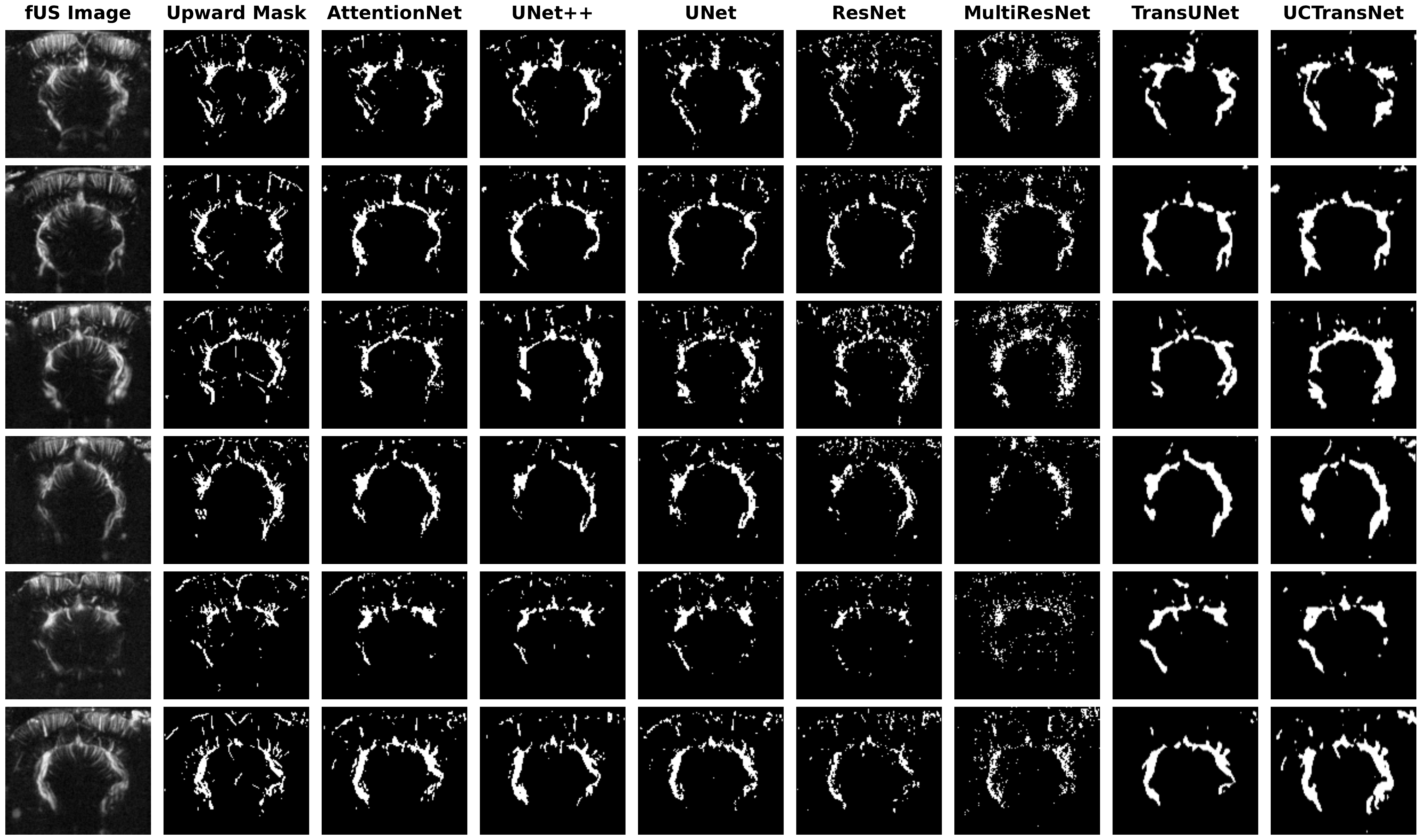}
\caption{Upward prediction examples using $\mathscr{L}_{CF}$. Each row depicts a different image, starting with the first frame of the fUS stack in log scale, followed by the constructed upward mask from ULM, and the segmentation proposed by the five benchmarked models.}
\label{fig:prediction_vein}
\end{figure}

For illustration, we presented predictions for six images in Figures~\ref{fig:prediction_artery} and \ref{fig:prediction_vein} for downward and upward flow, respectively. These figures are complemented by error images (false positives and negatives) in Appendix. B.
Surprisingly, despite achieving the highest IoU and F1 scores, the predictions made by TransUNet are imprecise. The resulting segmentations appear as amorphous blobs covering the relevant areas, lacking the distinct characteristics inherent to vascular structures. A similar observation applies to UCTransNet. This suggests that these models may be better suited for segmenting organs, tumors, or other convex and well-defined shapes rather than vascular patterns.
The predictions from Attention UNet and UNet++ align well with the performance metrics, closely matching the ground truth in the six examples. However, they struggle slightly with fine arterial structures at the cortical level and horizontal vessels, as seen in example 4. UNet and ResNet predictions respect the overall shape of the six images, but there are many missing structures, indicating that these models may not capture all the intricate details effectively. 
Finally, the segmentation of MultiresNet is notably imprecise, yet it shows slightly better performance in segmenting upward compartments that have a simpler structure. 

In the above experiment, we used the full fUS stack of 3 000 frames as input for training. We then repeated the experiment, this time using a single image per fUS stack, generated by averaging the 3 000 frames captured at rest. No statistically significant difference in overall performance was observed. Detailed results are provided in Appendix. C.
This additional experiment was not conducted with TransUNet and UCTransNet due to the high computational cost of training these models and their inability to accurately capture the fine structure of blood vessels.

\begin{table}[t]
\small
\centering
\caption{Performance comparison of models and loss functions trained on full fUS stacks at rest. Bold values represent the best performance on average for each model across the loss functions. Statistical significance tests between model-loss pairs are detailed in~\ref{sec:stat_tests_3000}.}
\setlength{\tabcolsep}{3pt}
\begin{tabular}{lcccccc}
\hline
\multicolumn{7}{c}{\textbf{Attention UNet}}\\ \hline
\textbf{Loss} & \textbf{Accuracy} $\uparrow$  & \textbf{F1 Score} $\uparrow$  & \textbf{Precision} $\uparrow$  & \textbf{Recall} $\uparrow$ & \textbf{Jaccard Index} $\uparrow$ & \textbf{Specificity} $\uparrow$ \\
\hline
$\mathscr{L}_{Dice\_CE}$ & $0.89 \pm 0.01$ & $0.67 \pm 0.02$ & $0.70 \pm 0.02$ & $0.65 \pm 0.03$ & $0.55 \pm 0.02$ & $0.86 \pm 0.02$  \\
$\mathscr{L}_{CF\_B}$  & $0.89 \pm 0.01$ & $0.67 \pm 0.02$ & $0.70 \pm 0.02$ & $0.65 \pm 0.03$ & $0.54 \pm 0.02$ & $0.85 \pm 0.01$  \\
$\mathscr{L}_{CF\_V}$  & $0.89 \pm 0.01$ & $0.68 \pm 0.01$ & $0.70 \pm 0.03$ & $0.67 \pm 0.02$ & $0.55 \pm 0.01$ & $0.86 \pm 0.01$ \\
$\mathscr{L}_{CF}$  & \textbf{0.89 $\pm$ 0.01} & \textbf{0.69 $\pm$ 0.01} & \textbf{0.71 $\pm$ 0.02} & \textbf{0.68 $\pm$ 0.02} & \textbf{0.56 $\pm$ 0.01} & \textbf{0.86 $\pm$ 0.01} \\
\hline

\multicolumn{7}{c}{\textbf{UNet++}}\\ \hline

$\mathscr{L}_{Dice\_CE}$  & $0.89 \pm 0.01$ & $0.67 \pm 0.01$ & $0.69 \pm 0.01$ & $0.66 \pm 0.02$ & $0.54 \pm 0.01$ & $0.86 \pm 0.01$  \\
$\mathscr{L}_{CF\_B}$ & $0.88 \pm 0.01$ & $0.67 \pm 0.01$ & $0.68 \pm 0.02$ & $0.66 \pm 0.01$ & $0.54 \pm 0.01$ & $0.86 \pm 0.00$  \\
$\mathscr{L}_{CF\_V}$  & $0.89 \pm 0.01$ & $0.67 \pm 0.01$ & $0.70 \pm 0.02$ & $0.67 \pm 0.02$ & $0.56 \pm 0.01$ & $0.87 \pm 0.01$ \\
$\mathscr{L}_{CF}$ & \textbf{0.89 $\pm$ 0.01} & \textbf{0.69 $\pm$ 0.01} & \textbf{0.71 $\pm$ 0.01} & \textbf{0.68 $\pm$ 0.02} & \textbf{0.56 $\pm$ 0.01} & \textbf{0.87 $\pm$ 0.01} \\
\hline

\multicolumn{7}{c}{\textbf{ResNet}}\\ \hline

$\mathscr{L}_{Dice\_CE}$ & $0.88 \pm 0.01$ & $0.63 \pm 0.01$ & \textbf{0.68 $\pm$ 0.02} & $0.61 \pm 0.02$ & $0.51 \pm 0.01$ & \textbf{0.84 $\pm$ 0.01}  \\
$\mathscr{L}_{CF\_B}$ & \textbf{0.88 $\pm$ 0.01} & \textbf{0.64 $\pm$ 0.02} & $0.67 \pm 0.03$ & \textbf{0.62 $\pm$ 0.02} & \textbf{0.51 $\pm$ 0.01} & $0.83 \pm 0.01$  \\
$\mathscr{L}_{CF\_V}$ & $0.88 \pm 0.01$ & $0.63 \pm 0.01$ & $0.67 \pm 0.02$ & $0.61 \pm 0.02$ & $0.51 \pm 0.01$ & $0.84 \pm 0.02$ \\
$\mathscr{L}_{CF}$ & $0.87 \pm 0.01$ & $0.61 \pm 0.01$ & $0.65 \pm 0.02$ & $0.59 \pm 0.03$ & $0.49 \pm 0.01$ & $0.84 \pm 0.02$ \\
\hline

\multicolumn{7}{c}{\textbf{UNet}}\\ \hline

$\mathscr{L}_{Dice\_CE}$ & $0.88 \pm 0.01$ & $0.66 \pm 0.01$ & \textbf{0.69 $\pm$ 0.02} & $0.64 \pm 0.02$ & $0.53 \pm 0.01$ & $0.85 \pm 0.01$  \\
$\mathscr{L}_{CF\_B}$ & $0.88 \pm 0.01$ & $0.65 \pm 0.01$ & $0.67 \pm 0.02$ & $0.64 \pm 0.01$ & $0.52 \pm 0.01$ & $0.85 \pm 0.01$  \\
$\mathscr{L}_{CF\_V}$  & $0.88 \pm 0.01$ & $0.67 \pm 0.02$ & $0.69 \pm 0.03$ & $0.66 \pm 0.02$ & $0.54 \pm 0.02$ & $0.86 \pm 0.01$ \\
$\mathscr{L}_{CF}$ & \textbf{0.88 $\pm$ 0.01} & \textbf{0.67 $\pm$ 0.01} & $0.68 \pm 0.02$ & \textbf{0.67 $\pm$ 0.02} & \textbf{0.54 $\pm$ 0.01} & \textbf{0.86 $\pm$ 0.01} \\
\hline

\multicolumn{7}{c}{\textbf{MultiresNet}}\\ \hline

$\mathscr{L}_{Dice\_CE}$ & $0.84 \pm 0.02$ & $0.57 \pm 0.02$ & $0.57 \pm 0.02$ & $0.57 \pm 0.01$ & $0.45 \pm 0.02$ & $0.82 \pm 0.01$  \\
$\mathscr{L}_{CF\_B}$ & \textbf{0.84 $\pm$ 0.01} & \textbf{0.57 $\pm$ 0.02} & \textbf{0.57 $\pm$ 0.03} & \textbf{0.59 $\pm$ 0.02} & \textbf{0.45 $\pm$ 0.02} & \textbf{0.84 $\pm$ 0.01}\\
$\mathscr{L}_{CF\_V}$ & $0.83 \pm 0.01$ & $0.56 \pm 0.02$ & $0.57 \pm 0.01$ & $0.57 \pm 0.03$ & $0.44 \pm 0.01$ & $0.83 \pm 0.03$  \\
$\mathscr{L}_{CF}$ & $0.83 \pm 0.01$ & $0.56 \pm 0.02$ & $0.55 \pm 0.02$ & $0.58 \pm 0.03$ & $0.44 \pm 0.02$ & $0.83\pm 0.02$  \\
\hline

\multicolumn{7}{c}{\textbf{UCTransNet}}\\ \hline

$\mathscr{L}_{Dice\_CE}$ \ & $0.87 \pm 0.02$ & $0.68 \pm 0.09$ & $0.72 \pm 0.04$ & $0.66 \pm 0.12$ & $0.54 \pm 0.08$ & \textbf{0.88 $\pm$ 0.06}  \\
$\mathscr{L}_{CF\_B}$ & \textbf{0.87 $\pm$ 0.01} & $0.68 \pm 0.04$ & 0.73 $\pm$ 0.04 & 0.66 $\pm$ 0.05 & 0.55 $\pm$ 0.04 & 0.86 $\pm$ 0.03\\
$\mathscr{L}_{CF\_V}$ & $0.87 \pm 0.01$ & $0.68 \pm 0.03$ & \textbf{0.74 $\pm$ 0.03} & $0.65 \pm 0.05$ & $0.55 \pm 0.03$ & $0.85 \pm 0.02$  \\
$\mathscr{L}_{CF}$ & $0.86 \pm 0.03$ & \textbf{0.68 $\pm$ 0.02} & $0.71 \pm 0.05$ & \textbf{0.67 $\pm$ 0.03} & \textbf{0.55 $\pm$ 0.03} & 0.87 $\pm$ 0.02  \\

\hline

\multicolumn{7}{c}{\textbf{TransUNet}}\\ \hline

$\mathscr{L}_{Dice\_CE}$  & $0.87 \pm 0.01$ & \textbf{0.70 $\pm$ 0.01} & $0.72 \pm 0.04$ & \textbf{0.69 $\pm$ 0.3} & \textbf{0.57 $\pm$ 0.01} & \textbf{0.87 $\pm$ 0.02}  \\
$\mathscr{L}_{CF\_B}$ & \textbf{0.88 $\pm$ 0.01} & $0.70 \pm 0.02$ & \textbf{0.73 $\pm$ 0.02} & 0.67 $\pm$ 0.02 & 0.57 $\pm$ 0.02 & 0.86 $\pm$ 0.01\\
$\mathscr{L}_{CF\_V}$& $0.87 \pm 0.01$ & 0.69 $\pm$ 0.02 & 0.72 $\pm$ 0.03 & $0.67 \pm 0.02$ & $0.56 \pm 0.02$ & $0.86 \pm 0.01$  \\
$\mathscr{L}_{CF}$ & $0.87 \pm 0.01$ & 0.68 $\pm$ 0.02 & $0.72 \pm 0.03$ & 0.66 $\pm$ 0.03 & 0.55 $\pm$ 0.02 & 0.86 $\pm$ 0.02  \\
\hline

\end{tabular}
\label{tab:comprehensive_model_performance_3000}
\end{table}

\subsubsection{Evaluating the Impact of fUS stack depth on segmentation performance}
\label{sec:exp:stackdepth}

\begin{figure}[t]
\centering%
\includegraphics[width=0.45\linewidth]{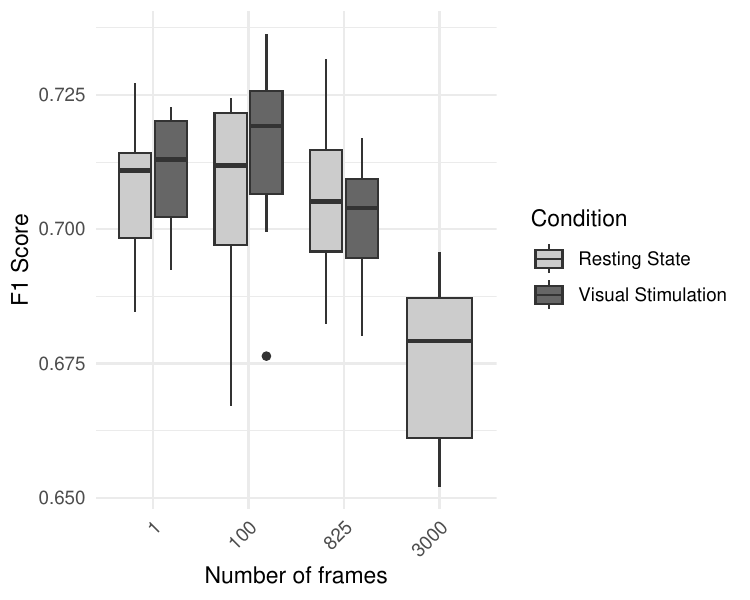}
\includegraphics[width=0.45\linewidth]{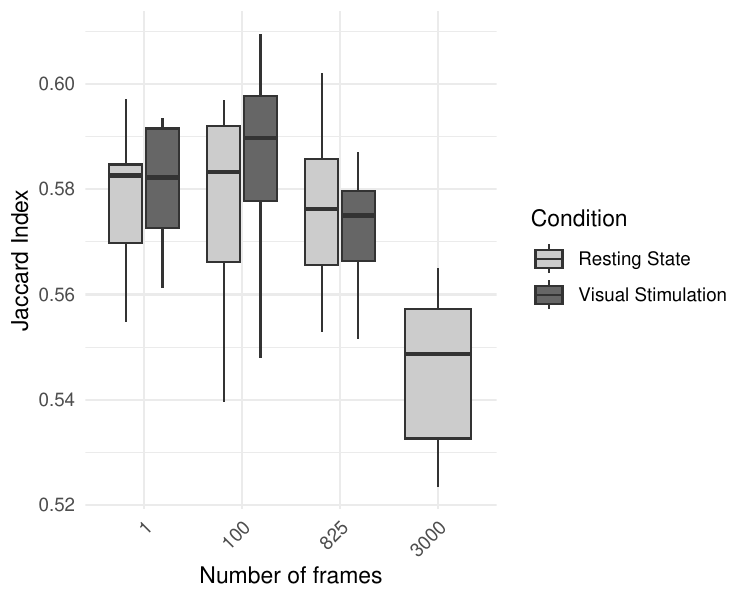}
\caption{Box plots illustrating the F1 score and Jaccard Index for predictions made by an Attention UNet trained on fUS stacks of varying depths ($1$, $100$, $825$, $3\,000$), both in resting state and under visual stimulation. A Wilcoxon paired test showed a statistically significant difference between the depths of $1$, $100$, $825$ and $3\,000$, whereas no significant difference is observed between the depths of $1$, $100$, $825$.} 
\label{fig:fus_depth_investigation}
\end{figure}

The primary goal of this experiment was to investigate how the depth of the fUS stack, defined by the number of frames used during training, impacts prediction quality. To address this question, we focused on the Attention UNet, one of the top-performing models from previous experiments, and trained it using $\mathscr{L}_{CF}$ on fUS stacks with varying numbers of frames. This analysis was conducted both at rest and under visual stimulation conditions, given the different signal evolution of cerebral blood volume observed in each state,  which can reach up to 10\%.
We worked with a total of 35 rats. For the resting-state condition, we trained the model on 30 rats, varying the number of frames used, and tested it on the remaining 5 rats. Similarly, for the visual stimulation condition, we re-trained the model from scratch on 30 rats, also varying the number of frames used, and tested it on the remaining 5 rats. These two experiments were conducted independently and the data were not mixed between the two conditions.
Furthermore, we emphasize that the reported results represent the average of a 7-fold cross-validation.
This approach ensured that every possible combination of training and testing subsets was tested, providing robust and comprehensive evaluation across all data.
We maintained the same sampling frequency for the stacks, varying only the number of frames, $n$, selected from the sets $\{1, 100, 825, 3\,000\}$ for the resting state and $\{1, 100, 825\}$ for visual stimulation. If the number of selected frames, $n$, was smaller than the total available frames, $N$ (i.e., $3\,000$ for resting state or $825$ for visual stimulation), we randomly selected $n$ contiguous frames from the full stack. Specifically, a random index $i$ was drawn from the distribution $i \sim \textsf{Uniform}(1, N - n)$  and the $i + n$ frames were used. This random selection was performed at every training step, meaning that different frames were sampled from each stack in every epoch of the training process.
The results for the F1 score and Jaccard Index are reported in Figure~\ref{fig:fus_depth_investigation}. 

These results show that both metrics --~F1 score and Jaccard Index~-- exhibited consistent trends across training scenarios in resting state and under visual stimulation, suggesting that signal evolution does not significantly affect segmentation quality. In particular, reducing the number of frames remarkably improved these metrics, achieving maximal Jaccard Index of $0.60$ and an F1 score above~$0.72$. A paired Wilcoxon test confirmed these improvements as statistically significant ($\text{p-value} < 0.05$) for the depths $1$, $100$ or $825$ compared to $3\,000$. This indicates that the model becomes easier to train with less parameters. 
Among the frame sets tested --~$1$, $100$, and $825$~-- the boxplot using $100$ frames ranks highest, particularly under visual stimulation.
This suggests that a single frame lacks adequate information and using $825$~frames might overly complicate the model, reducing its effectiveness. Thus, it appears that using $100$ frames provides sufficient data for accurate segmentation without overburdening the model.

\subsubsection{Cross-condition efficacy of fUS-based models: Training on resting state for visual stimulation segmentation}
\label{sec:exp:crosscondition}
We found above that brain signal changes captured by fUS during visual stimulation did not significantly influence the quality of segmentation, as similar results were achieved with or without visual stimulation. This led us to explore whether a model trained on fUS data from a resting state could effectively segment images acquired during visual stimulation. 
To this end, we trained the Attention UNet model on fUS images at rest (with $100$~frames) and tested it on images from visually stimulated sessions (also with $100$~frames). To ensure a clear separation between training and testing data, we used different groups of rats for each condition: fUS images from resting state sessions were used for training, while only fUS images from visual stimulation sessions of separate rats were used for testing. The goal was to ensure that the model was evaluated on new structural data.

\begin{table}[tb]
\centering
\caption{Performance of Attention UNet trained, with a $CF$ Loss, on fUS stacks ($100$~frames) in resting state and tested on fUS stacks ($100$~frames) under visual stimulation.}
\setlength{\tabcolsep}{2pt}
\begin{tabular}{lcccccc}
\hline
Model & Accuracy $\uparrow$  & F1 Score $\uparrow$  & Precision $\uparrow$  & Recall $\uparrow$ & Jaccard Index $\uparrow$ & Specificity $\uparrow$ \\
\hline
Attention-UNet & $0.90 \pm 0.00$ & $0.71 \pm 0.01$ & $0.73 \pm 0.02$ & $0.70\pm 0.02$ & $0.59 \pm 0.01$ & $0.89 \pm 0.01$  \\
\hline
\end{tabular}
\label{tab:training_rest_testing_vis_stim}
\end{table}

The results for various metrics are presented in Table~\ref{tab:training_rest_testing_vis_stim}, revealing an accuracy of $90\%$, an F1 score of $71\%$, and a Jaccard Index of $0.59$. These outcomes are noteworthy as they align with previous results. This is particularly interesting because it shows that our model can be effectively trained on shorter fUS stacks in a resting state, and still perform well on data collected during visual stimulation without requiring retraining for the new conditions. This highlights the model applicability for accurately tracking differentiated activity in different vascular compartments.

\subsubsection{Visualizing vascular structures in fUS imaging}
\label{sec:exp:visu}

In this section, we illustrated that the UNet based inference of 
upward and downward masks can be effectively used to enhance the interpretation of fUS signals. 

\begin{figure}[t]
\centering%
\includegraphics[width=\linewidth]{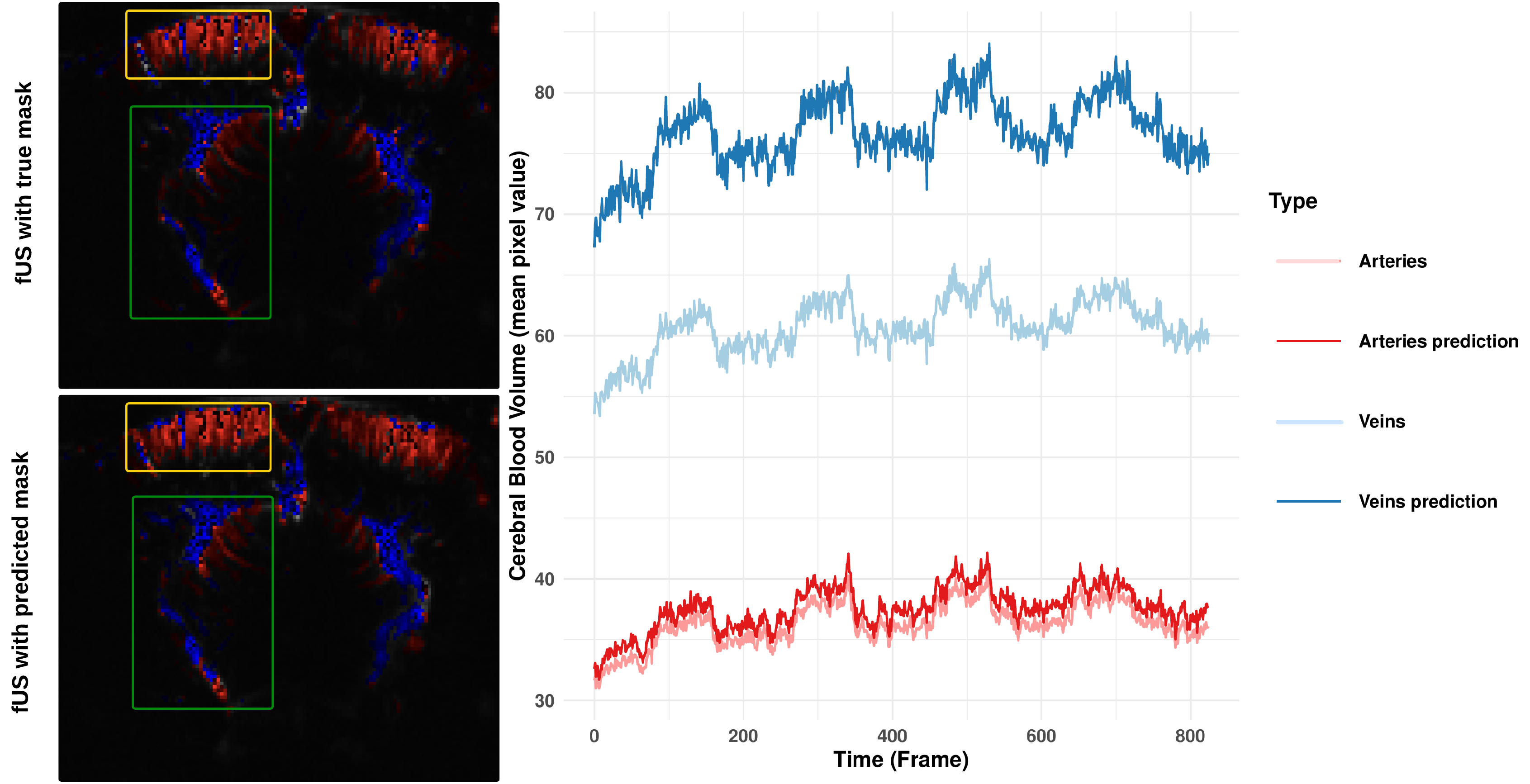}
\caption{Images on the left show the projection of real masks of upward and downward flow compared to predicted ones on a random frame taken from a fUS stack acquired under visual stimulation. Upward flow is depicted in blue and downward in red, with color intensity reflecting the cerebral signal strength during visual stimulation. Green rectangles highlight regions predicted with relative accuracy, while yellow rectangles indicate areas where the model struggled more.
The figure to right shows the temporal evolution of the mean signal intensity extracted from the cortical region of the same stack of fUS, separating signal from pixels classified as veins or arteries and displaying the predictions.} 
\label{fig:fus_with_masks}
\end{figure}

Firstly, for a more visual rendering of our results, we overlaid real and model-predicted masks on an fUS image captured during visual stimulation (Figure~\ref{fig:fus_with_masks}, left). Arterial and venous signals are distinguished by red and blue colors, respectively, while the background is rendered in grayscale. The color intensity corresponds to the strength of the Power Doppler signal, indicative of Cerebral Blood Volume (CBV). Figure~\ref{fig:fus_with_masks} illustrates the outcome of this visualization process.
We surrounded regions where the model predictions were relatively accurate in green and areas where it struggled in yellow. 
It should be noted that the cortex is located at the top of the images. 
In the cortical region, the predicted mask accurately delineates arteries but does not capture all the veins, although it does detect a fair number of them. This difference in performance may be due to the denser presence of arteries compared to veins in this region. In contrast, the model performs particularly well in the lower regions, where the signal appears more visible, and it effectively identifies both downward and upward compartments.

This visualization was complemented by extracting the signal evolution in the cortical region across different compartments within the same fUS stack. Specifically, we averaged the pixels classified as veins or arteries per frame in the cortex, allowing us to visualize the evolution of the CBV in the graph to the right of Figure~\ref{fig:fus_with_masks}.
Initially, we can distinguish four patterns corresponding to the response to four sessions of visual stimulation. The signals from both predicted and actual arteries are almost identical, achieving a Pearson correlation coefficient of $0.99$. However, a difference in intensity between the signals from predicted and actual veins is noticeable. This discrepancy may be explained by the model inability to accurately identify veins in the cortical region due to their lower density in fUS images. This is consistent with the illustration showing the model limitations on the left. Aside from this, the signal follows the same trend, and the Pearson correlation coefficient is $0.98$, indicating a strong linear correlation. We calculated the average correlation coefficients across all rats and achieved values of $0.98$ for arteries and $0.55$ for veins in the cortical region and $0.87$ for downward flow and $0.98$ for upward flow in the lower region.

In conclusion, while the model effectively identifies arterial structures within fUS imaging, it struggles more with accurately capturing venous structures in the cortical region. This limitation highlights the need for model enhancements to improve detection capabilities for less dense vascular features. Nonetheless, the correlation remains strong in both compartments, underscoring the model overall utility.

\section{Conclusion and Perspectives}
\label{sec5}

In this article, we introduced the first study on blood flow segmentation (upward vs. downward) of the brain using fUS imaging. Annotations were automatically derived from costly ultrasound localization microscopy, which is no longer required after model training to detect upward and downward flows. We evaluated several segmentation models, originally developed for other imaging modalities, and achieved competitive performance comparable to existing vascular analysis methods. Notably, using only $100$~frames of the fUS stack during training yielded optimal results, with $90\%$ accuracy and an F1 score of $71\%$.
Furthermore, our findings showed that training the model on fUS data acquired in a resting state is sufficient for accurately segmenting images obtained during visual stimulation sessions. This demonstrates the model potential for analyzing brain activity during task-based experiments. The predicted masks also provided significant value in enhancing the interpretation of fUS Power Doppler signals, which is particularly useful for neuroscientists conducting preclinical studies. Our pipeline demonstrated a high level of agreement between predicted and actual signals, with average linear correlation coefficients of $0.98$ for arteries and $0.55$ for veins in the cortical region, and $0.87$ for downward and $0.98$ for upward flows in deeper regions. 
Although the model effectively captures arterial structures, it faces challenges with venous segmentation in the cortical region, highlighting potential areas for improvement. Nevertheless, the consistently high correlation underscores the model robustness and overall reliability.
However, the model generalization to other brain regions requires staying within the same imaging plane, meaning a new model must be trained for each plane with ULM data. While this introduces certain constraints, the ongoing development of 3D fUS and ULM technologies is expected to mitigate this limitation in the near future.
We emphasize that our annotation framework is reliable in the cortical region, where blood vessels exhibit a predominantly vertical orientation, but it has limitations in subcortical regions. Developing a more robust annotation method, potentially integrating additional directional flow information or anatomical priors, could significantly enhance the accuracy of ground truth generation. This represents a key direction we intend to pursue in future work.
As a proof of concept, this work lays the foundation for further advances. Future improvements could include partitioning fUS images into patches and training the model to segment regions independently of their location. Additionally, a better use of the temporal dimension of fUS stacks could be achieved by transforming them into a suitable latent space using advanced techniques like the Inception Time model~\cite{Inceptiontime} to capture signal variations more effectively. Another promising direction is to develop a UNet architecture specifically tailored to the unique signal characteristics of fUS, particularly improving its sensitivity to less dense vascular structures, or to leverage advanced models such as the Dual Multi-Scale Attention UNet (DMSA-UNet)~\cite{DMSA-UNet}, which has shown strong potential in medical image segmentation.

\section*{Acknowledgements}
This work was funded by the ANR (LabCom “NI2D” ANR-20-LCV1-0001) and the Auvergne-Rhône-Alpes Region (Pack Ambition Recherche “Brain Imaging for Drug Discovery”).

\section*{Ethics Statement}
All animal experimentation and housing were carried out in accordance with the guidelines of the European Union (Directive 2010/63/EU) and approved by the French Ministry of Research (authorization reference: APAFIS-36043). The animals were kept under controlled environmental conditions (12/12 h light-dark cycle, food and water ab libitum, controlled temperature) with at least one week of acclimation before any experiment. All efforts were made to ensure the animal welfare. This dataset was re-used from a previous study, avoiding the use of new animals \cite{Pereira2024}.

\bibliographystyle{elsarticle-num}

\newpage

\appendix

\section{Statistical Significance of Model Comparisons (trained on full fUS stacks) Using the Wilcoxon Test}

\label{sec:stat_tests_3000}
\scriptsize
\begin{center}
\begin{longtable}{lllllll}
\caption{P-values from the paired Wilcoxon significance test for the different metrics from the training on the full fUS stacks (values lower than 0.05, highlighted in bold in the table, indicate that the observed differences are significant)}
\\
  \hline
Comparison & F1Score & Accuracy & Precision & Recall & IoU & Specificity \\ 
\hline
(AttentionUNet-DICE\_CE) vs (MultiresNet-DICE\_CE) & \textbf{0.016} & \textbf{0.016} & \textbf{0.016} & \textbf{0.016} & \textbf{0.016} & \textbf{0.016} \\ 
  (AttentionUNet-DICE\_CE) vs (UNet++-DICE\_CE) & 0.47 & 0.58 & \textbf{0.047} & 0.94 & 0.47 & 0.94 \\ 
  (AttentionUNet-DICE\_CE) vs (ResNet-DICE\_CE) & \textbf{0.016} & \textbf{0.047} & 0.22 & \textbf{0.016} & \textbf{0.016} & 0.078 \\ 
  (AttentionUNet-DICE\_CE) vs (UNet-DICE\_CE) & 0.3 & 0.3 & 0.38 & 0.47 & 0.22 & 0.22 \\ 
  (AttentionUNet-DICE\_CE) vs (UCTransNet-DICE\_CE) & 0.94 & \textbf{0.016} & 0.3 & 0.58 & 1 & 0.81 \\ 
  (AttentionUNet-DICE\_CE) vs (TransUNet-DICE\_CE) & \textbf{0.016} & \textbf{0.031} & 0.38 & \textbf{0.047} & 0.078 & 0.078 \\
\hline

\end{longtable}
\end{center}

\section{Error Images for Model Comparisons (trained on full fUS stacks)}

\begin{figure}[t]
\centering%
\includegraphics[width=\linewidth]{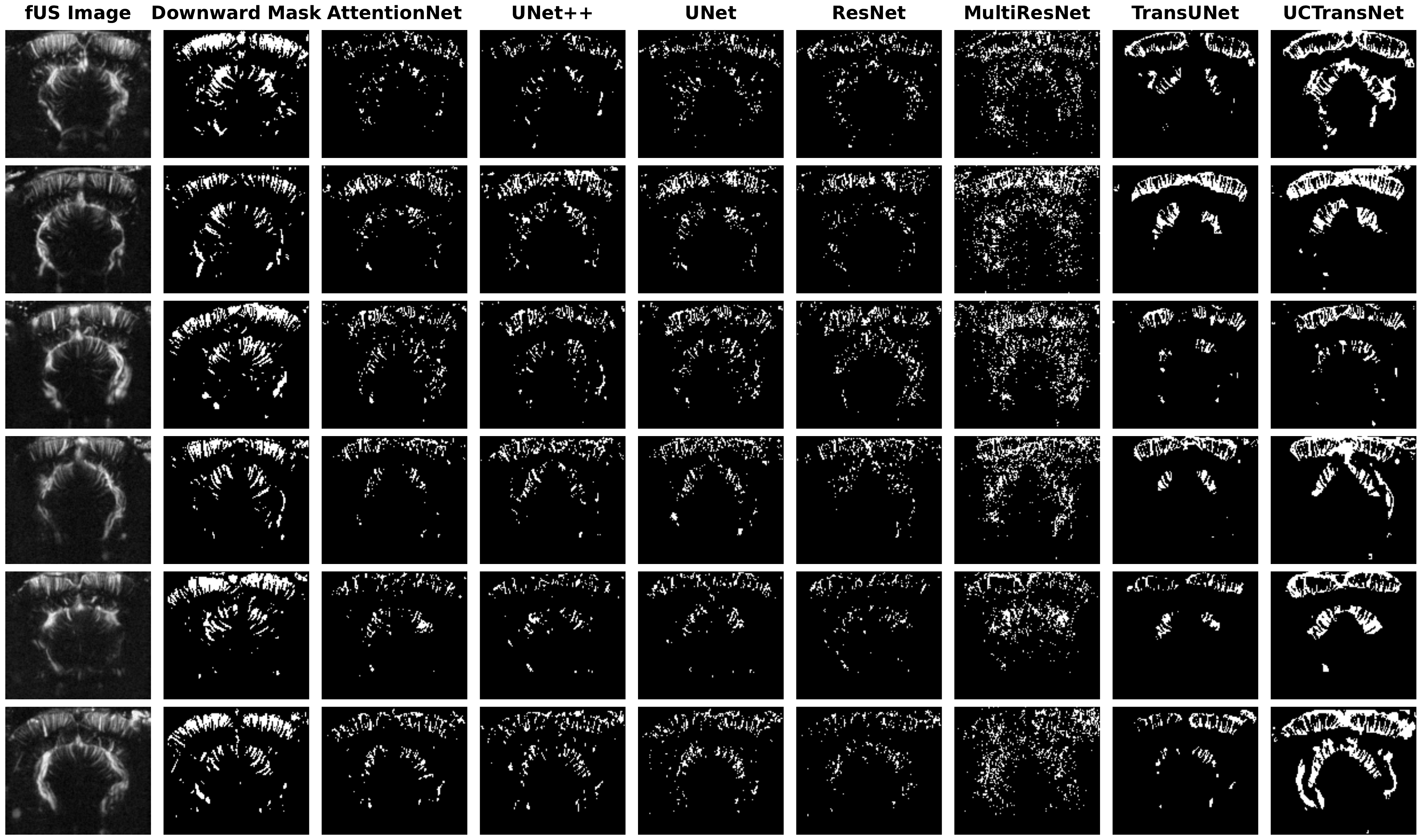}
\caption{Downward error predictions (false positives) examples using $\mathscr{L}_{CF}$. Each row depicts a different image, starting with the first frame of the fUS stack in a log scale, followed by the constructed downward mask from ULM, and the errors of the five benchmarked models}
\label{fig:prediction_artery_fp}
\end{figure}

\begin{figure}[t]
\centering%
\includegraphics[width=\linewidth]{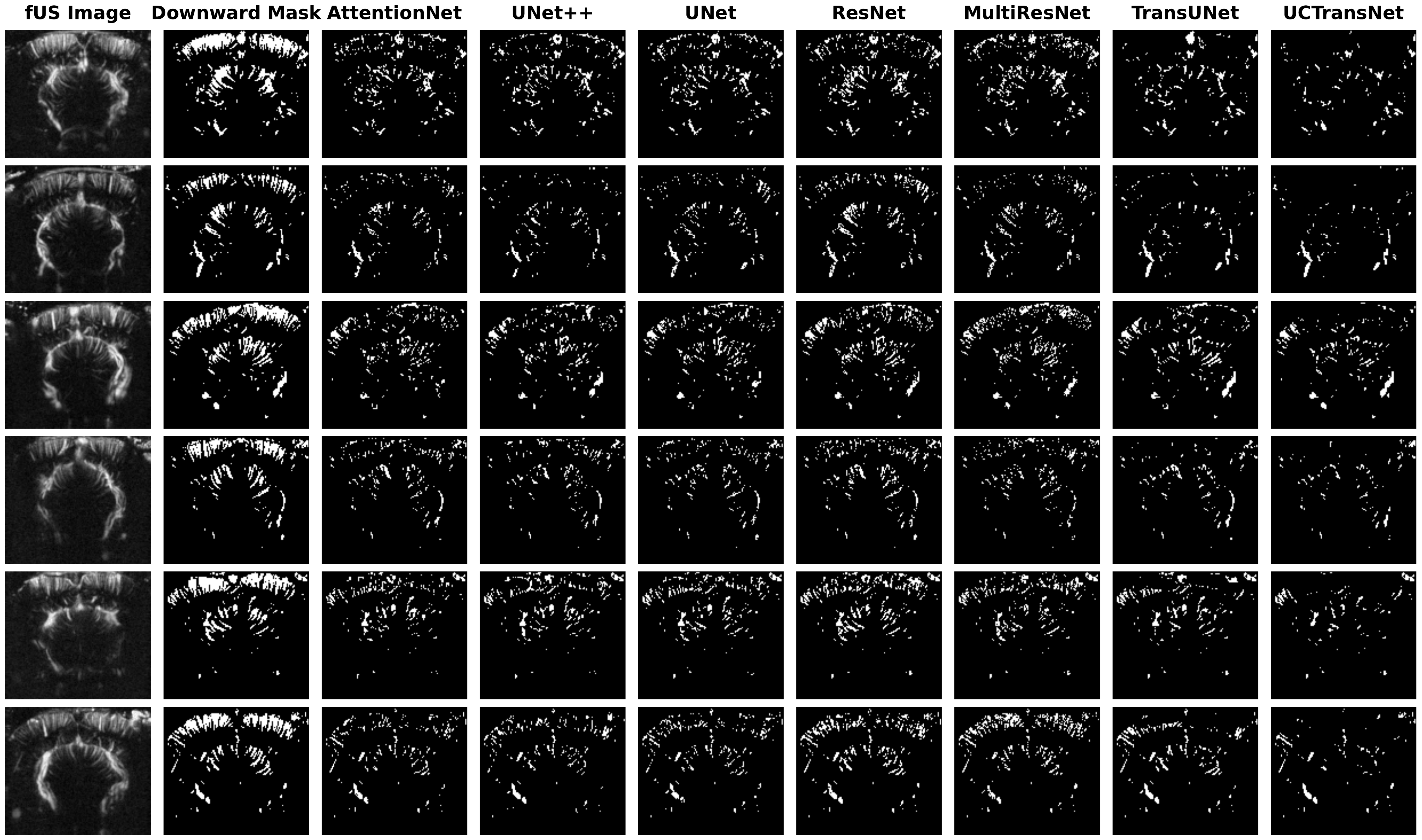}
\caption{Downward error predictions (false negatives) examples using $\mathscr{L}_{CF}$. Each row depicts a different image, starting with the first frame of the fUS stack in a log scale, followed by the constructed downward mask from ULM, and the errors of the five benchmarked models}
\label{fig:prediction_artery_fn}
\end{figure}

\begin{figure}[t]
\centering%
\includegraphics[width=\linewidth]{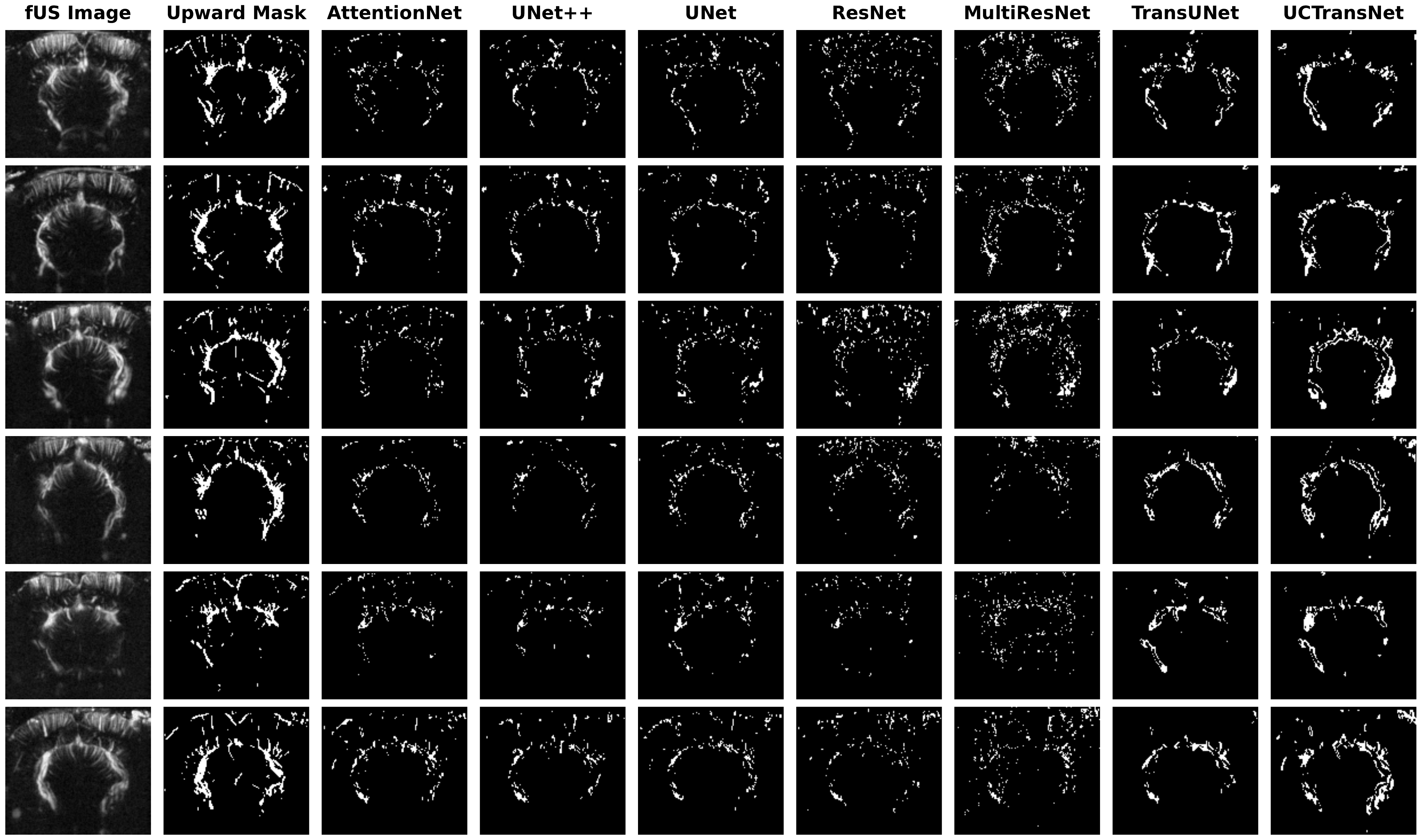}
\caption{Upward error predictions (false positives) examples using $\mathscr{L}_{CF}$. Each row depicts a different image, starting with the first frame of the fUS stack in a log scale, followed by the constructed upward mask from ULM, and the errors of the five benchmarked models}
\label{fig:prediction_vein_fp}
\end{figure}

\begin{figure}[t]
\centering%
\includegraphics[width=\linewidth]{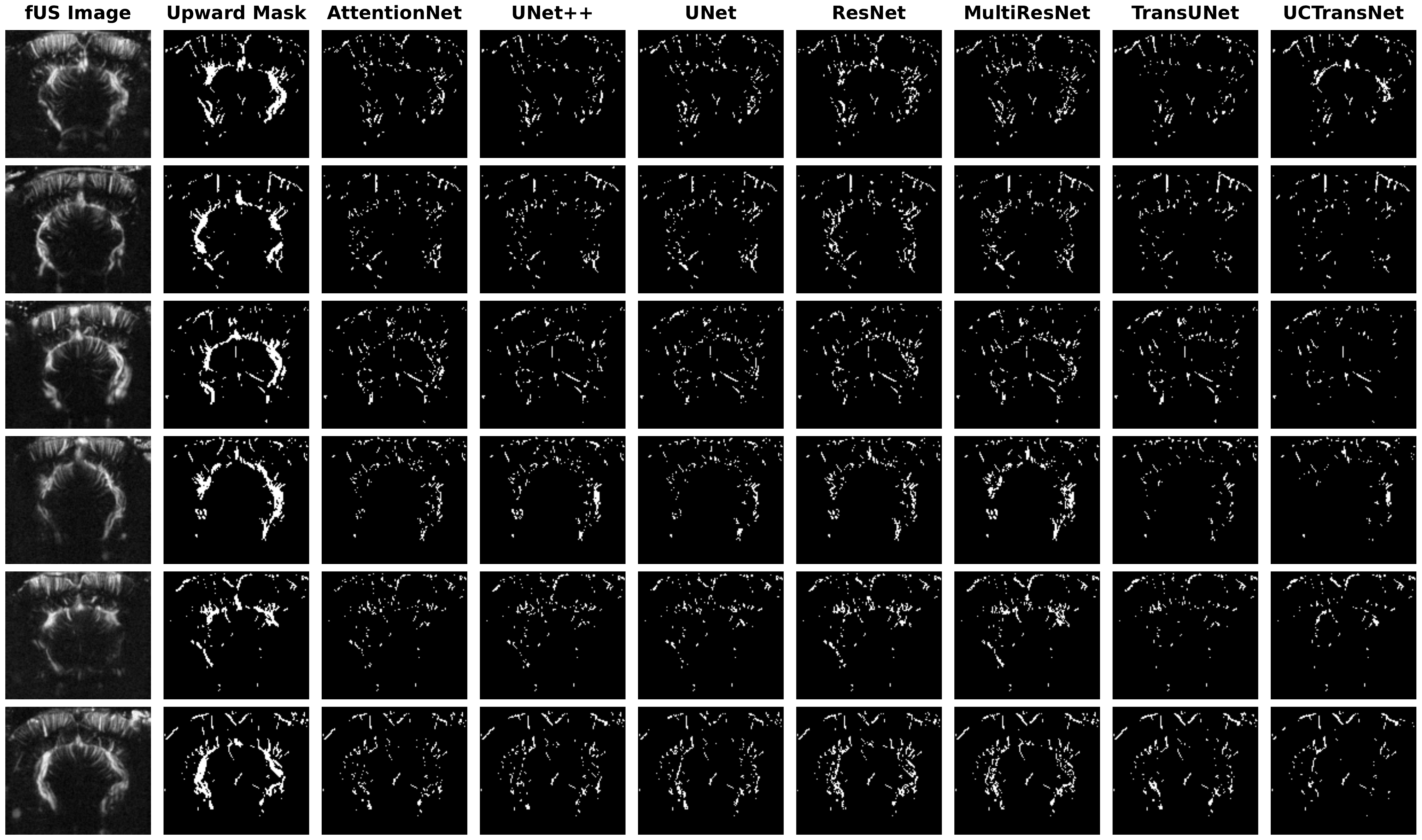}
\caption{Upward error predictions (false negatives) examples using $\mathscr{L}_{CF}$. Each row depicts a different image, starting with the first frame of the fUS stack in a log scale, followed by the constructed upward mask from ULM, and the errors of the five benchmarked models}
\end{figure}

% //

\section{Comparison among models and losses trained on averaged fUS stacks}

\begin{table}[t]
\small
\centering
\caption{Performance comparison across models and losses trained on averaged fUS stacks. Bold values represent the best performance for each model across the loss functions. Statistical significance tests between model-loss pairs are detailed in \ref{stat_tests_average}.}
\begin{tabular}{lcccccc}
\hline
\multicolumn{7}{c}{\textbf{Attention UNet}}\\ \hline
\textbf{Loss} & \textbf{Accuracy} & \textbf{F1 Score} & \textbf{Precision} & \textbf{Recall} & \textbf{Jaccard Index} & \textbf{Specificity} \\
\hline
$\mathscr{L}_{Dice\_CE}$ & $0.88 \pm 0.01$ & $0.67 \pm 0.02$ & $0.70 \pm 0.02$ & $0.65 \pm 0.02$ & $0.54 \pm 0.01$ & $0.85 \pm 0.01$  \\
$\mathscr{L}_{CF\_B}$  & $0.89 \pm 0.01$ & $0.67 \pm 0.02$ & $0.69 \pm 0.02$ & $0.65 \pm 0.02$ & $0.54 \pm 0.02$ & $0.85 \pm 0.01$  \\
$\mathscr{L}_{CF\_V}$  & $0.89 \pm 0.01$ & $0.68 \pm 0.01$ & \textbf{0.71 $\pm$ 0.02} & 0.67 $\pm$ 0.02 &  $0.55 \pm 0.01$ & $0.86 \pm 0.01$ \\
$\mathscr{L}_{CF}$ & \textbf{0.89 $\pm$ 0.01} & \textbf{0.68 $\pm$ 0.01} & $0.70 \pm 0.01$ & \textbf{0.68 $\pm$ 0.02} & \textbf{0.55 $\pm$ 0.01} & \textbf{0.86 $\pm$ 0.01}  \\
\hline

\multicolumn{7}{c}{\textbf{MultiresNet}}\\ \hline

%\textbf{Loss} & \textbf{Accuracy} & \textbf{F1 Score} & \textbf{Precision} & \textbf{Recall} & \textbf{Jaccard Index} & \textbf{Specificity} \\
%\hline
$\mathscr{L}_{Dice\_CE}$ & \textbf{0.84 $\pm$ 0.02} & \textbf{0.58 $\pm$ 0.02} & \textbf{0.57 $\pm$ 0.02} & \textbf{0.60 $\pm$ 0.01} & \textbf{0.46 $\pm$ 0.02} & \textbf{0.84 $\pm$ 0.01}  \\
\hline
\end{tabular}
\label{tab:comprehensive_model_performance_1}
\end{table}

\begin{center}
\scriptsize
\setlength{\tabcolsep}{4pt}
\begin{longtable}{lllllll}
\caption{P-values from the paired Wilcoxon significance test for the different metrics from the training on the averaged fUS stacks (values lower than 0.05, highlighted in bold in the table, indicate that the observed differences are significant)}
\\
  \hline
Comparison & F1Score & Accuracy & Precision & Recall & IoU & Specificity \\ 
  \hline
(AttentionUNet-DICE\_CE) vs (MultiresNet-DICE\_CE) & \textbf{0.016} & \textbf{0.016} & \textbf{0.016} & \textbf{0.016} & \textbf{0.016} & 0.11 \\ 
  (AttentionUNet-DICE\_CE) vs (UNet++-DICE\_CE) & 1 & 0.81 & 0.3 & 0.58 & 1 & 0.3 \\ 
  (AttentionUNet-DICE\_CE) vs (ResNet-DICE\_CE) & \textbf{0.016} & 0.11 & \textbf{0.016} & \textbf{0.016} & \textbf{0.016} & \textbf{0.016} \\ 
  (AttentionUNet-DICE\_CE) vs (UNet-DICE\_CE) & 0.22 & 0.16 & 0.47 & 0.3 & 0.22 & 0.47 \\ 
  (AttentionUNet-DICE\_CE) vs (AttentionUNet-CF\_B) & 0.94 & 1 & 1 & 1 & 0.81 & 1 \\ 
  (AttentionUNet-DICE\_CE) vs (MultiresNet-CF\_B) & \textbf{0.016} & \textbf{0.016} & \textbf{0.016} & \textbf{0.016} & \textbf{0.016} & \textbf{0.016} \\
\end{longtable}
\end{center}

\end{document}